\documentclass[preprint,showpacs,showkeys,preprintnumbers,amsmath,amssymb]{revtex4}
\usepackage{dcolumn,pstricks}
\usepackage{bm}

\numberwithin{equation}{section}

\renewcommand{\a}{\alpha}           
\renewcommand{\b}{\beta}            
\newcommand{\ga}{\gamma}            
\newcommand{\gm}{\Gamma}            
\newcommand{\dl}{\delta}

\newcommand{\ka}{\kappa}               
\newcommand{\la}{\lambda}           
           
\newcommand{\m}{\mu}             
\newcommand{\n}{\nu}             
\newcommand{\om}{\omega}          
\newcommand{\si}{\sigma}          
          
\renewcommand{\th}{\theta} 
\newcommand{\pa}{\partial}
\newcommand{\na}{\nabla}

\newcommand{\boldE}{\mathbf{E}}  
\newcommand{\boldB}{\mathbf{B}}

\newcommand{\hk}{\hat{k}}  
\newcommand{\hphi}{\hat{\varphi}}
\newcommand{\ho}{\hat{\om}} 
\newcommand{\bA}{\bar{A}}
\newcommand{\bB}{\bar{B}} 
\newcommand{\bE}{\bar{E}} 
\newcommand{\hH}{\hat{H}} 
\newcommand{\bF}{\bar{F}}
\newcommand{\hF}{\hat{F}} 
\newcommand{\bg}{\bar{g}} 
\newcommand{\hg}{\hat{g}}
\newcommand{\hK}{\hat{K}} 
\newcommand{\bR}{\bar{R}} 
\newcommand{\hR}{\hat{R}}
\newcommand{\bT}{\bar{T}} 
\newcommand{\hT}{\hat{T}}
\newcommand{\bna}{\,\overline{\!\nabla\!}\,}
\newcommand{\thbF}{\th\hspace{-1.5pt}\cdot\hspace{-2.2pt}\bF}
\newcommand{\thdbF}{\th\hspace{-1.5pt}\cdot\hspace{-2.2pt}
         \tilde{\bar{F}}}
\newcommand{\thbf}{\th\hspace{-1.5pt}\cdot\hspace{-2.2pt}\bar{f}}
\newcommand{\ds}{\displaystyle}

\begin{document}

\preprint{}

\title{Noncommutative Einstein-Maxwell \textit{pp\,}-waves}

\author{S. Marculescu}
\affiliation{Fachbereich Physik, Universit\"at Siegen, D-57068 Siegen, Germany}

\author{F. Ruiz Ruiz}
\affiliation{Departamento de F\'{\i}sica Te\'orica I, Universidad
Complutense de Madrid, 28040 Madrid, Spain}

\date{\today}

\begin{abstract}
 The field equations coupling a Seiberg-Witten electromagnetic field to
  noncommutative gravity, as described by a formal power series in the
  noncommutativity parameters $\theta^{\alpha\beta}$, is investigated.  A
  large family of solutions, up to order one in $\theta^{\alpha\beta}$, describing
  Einstein-Maxwell null {\it pp\,}-waves is obtained. The order-one
  contributions can be viewed as providing noncommutative corrections to {\it
    pp\,}-waves. In our solutions, noncommutativity enters the spacetime
  metric through a conformal factor and is responsible for
  dilating/contracting the separation between points in the same null surface.
  The noncommutative corrections to the electromagnetic waves, while
  preserving the wave null character, include constant polarization, higher
  harmonic generation and inhomogeneous susceptibility. As compared to pure
  noncommutative gravity, the novelty is that nonzero corrections to the
  metric already occur at order one in $\theta^{\alpha\beta}$.
\end{abstract}

\pacs{11.10.Nx, 04.40.Nr}

\keywords{Noncommutative corrections, {\it pp\,}-waves}

\maketitle

\section{\label{sec:Introduction}Introduction}

The idea that physics at the Planck length may be a probe for
noncommutativity~\cite{Doplicher} has made of noncommutative gravity a central
issue in the related literature; see ref.~\cite{Szabo} for a review.  By now,
several noncommutative deformations of general relativity in various
dimensions~\cite{grav-def} have been proposed with varied luck, many of which
involve complexification of the metric and of the local Lorentz group. Lately
a more fundamental theory of noncommutative gravity based on a deformation of
the group of diffeomorphisms~\cite{Munich} has been proposed, and its
connection with the Seiberg-Witten limit~\cite{Seiberg-Witten} of graviton
interactions in a bosonic string theory has also been studied~\cite{AG-VM}.
This formulation has the property, shared by many phenomenological
approaches~\cite{grav-def}, that an expansion in powers of the
noncommutativity parameters $\th^{\a\b}$ reveals that corrections to Einstein
gravity start at order two. However, no explicit solution to the corresponding
field equations has been found to this order.

In this paper, inspired by recent results on noncommutatively smeared
Schwarzschild black holes~\cite{black-hole}, we couple general
  relativity to a Seiberg-Witten electromagnetic (EM) field up to order one in
  $\th^{\a\b}$.  Our model is based on two assumptions.  The first one is
that Einstein gravity should remain applicable.  This is justifiable since, as
already mentioned, in more fundamental approaches~\cite{Munich,AG-VM} and in
many deformation approaches~\cite{grav-def,Mukherjee} noncommutative
corrections start at order two in $\th^{\a\b}$. The second hypothesis is that
noncommutative gravity modifications should already occur at order one in
$\th^{\a\b}$, since matter distributions from field theory classical actions
receive contributions to this order.  In other words, even though
  gravity lacks first order corrections in $\th^{\a\b}$, the right-hand side
  of the Einstein equations provides such corrections.  Based on these
  assumptions, we consider a model that couples gravity, described by the
  Einstein-Hilbert action, to the order-zero and order-one terms of the
  Seiberg-Witten expansion for the action of a gauge field. This yields the
  classical action in~\eqref{action} below, which defines our model. In this
  paper we will look at the corrections to the metric due to the occurrence of
  $\th^{\a\b}$ on the right-hand side of the Einstein equations, and to the
  $\th^{\a\b}$-dependence of the EM field.

Because of their relevance in general relativity and in string propagation on
gravitational backgrounds, we are interested here in finding {\it pp\,}-wave
solutions. We anticipate ourselves and mention that we are able to construct a
large variety of noncommutative null {\it pp\,}-wave spacetimes. In all of
them, the noncommutativity dependence on the metric is through a conformal
factor which is a function of the coordinate labeling the null surface of
spacetime. In turn the EM field receives in general three different types of
noncommutative corrections: a dynamically generated constant
polarization/induction contribution that can be tuned at will, a
susceptibility inhomogeneous polarization due to gravity, and a nonlinear
dipolar contribution, ultimately caused by the Seiberg-Witten map, which
generates higher harmonics. These EM waves do not exhibit modified dispersion
relations and are of a different type to those encountered in flat
spacetime~\cite{Guralnik}.

The paper is organized as follows. We introduce in Section II the classical
action defining our model and describe the symmetry principle behind it. In
Section III we derive the corresponding Einstein and EM field equations.  Since
$\th^{\a\b}$ is treated as a small parameter, the field equations decouple
into an order-zero system of equations, describing conventional
Einstein-Maxwell theory, and a complicated order-one system of equations. Also
in Section III we make a conformal ansatz for the noncommutative correction to
the metric that simplifies very much the order-one equations. We consider in
Section IV solutions of the order-zero problem describing null Einstein-Maxwell
{\it pp\,}-waves, and ask ourselves whether the order-one equations have
solutions preserving the {\it pp\,}-wave nature of the the such order-zero
backgrounds. The answer is in the affirmative and does not rely on the details
of the equations but rather on the conformal ansatz made for the
noncommutative correction to the metric. In Section V we move on to study
particular cases of noncommutative {\it pp\,}-wave spacetimes.  The physical
significance of the noncommutative corrections to the EM field and a
comparison with known solutions in the literature for flat spacetime is
performed in Section VI. We conclude in Section VII.

\section{\label{sec:action}Classical action}

Our interest is to study the noncommutative coupling of an Abelian gauge field
$A_\a$ to Einstein gravity to order one in the noncommutativity parameters
$\th^{\a\b}$. Let us first write the classical action that defines our model
and then discuss how it arises. The action is given by
\begin{equation}
   S = \int d^4\!x\, \sqrt{-g} \> \Big[\, \frac{R}{2\kappa^2}
      - \frac{1}{4}\, F_{\a\b}  F^{\a\b} 
      - \frac{1}{2}\, \th^{\m\n} \Big( F_{\m\a} F_{\n\b} 
          - \frac{1}{4}\, F_{\m\n} F_{\a\b} \Big) F^{\a\b} \Big] 
   + O(\th^2)\,,
\label{action}
\end{equation}
where $F_{\a\b}= \pa_\a A_\b - \pa_\b A_\a$ is the field strength. To explain
this action, we situate ourselves in a reference frame in which $\th^{\a\b}$ is
constant and invoke the two hypotheses mentioned in the introduction. The
first one was that general relativity should be applicable up to order in
$\th^{\a\b}$, since existing formulations~\cite{Munich} of noncommutative
gravity based on a deformation of the diffeomorphism group introduced
noncommutative corrections to general relativity at order two and higher.
Hence the gravity part of the classical action up to order one should be the
Einstein-Hilbert action, which is the first term in~\eqref{action}. The second
and third terms are the most straightforward generalization to curved
spacetime of the action provided in flat spacetime by the Seiberg-Witten map
for a $U(1)$ gauge field. These terms are in agreement with the second
assumption above. Next we adopt the observer point of view~\cite{Colladay} and
regard $\th^{\a\b}$ as a contravariant two-tensor. This implies that the
action~\eqref{action} is invariant under conventional gauge transformations of
$A_\a$ and diffeomorphisms. Note that, since we are not dealing with Moyal
products, we are not running into the problems associated with nonconstant
noncommutativity~\cite{Gayral-GB-Ruiz}. This defines our model and is the
starting point for our analysis.

This action can be viewed as collecting the order-zero and order-one terms of
a formal power series in $\th^{\a\b}$ for a classical action describing the
Seiberg-Witten coupling of gravity and electromagnetism. Our solutions must be
understood in this way, as providing the first nontrivial terms of a formal
power series in $\th^{\a\b}$ describing noncommutative deformations of both
gravitational and EM fields.  The lack of first order contributions to the
gravity sector of the action~\cite{grav-def,Munich,AG-VM,Mukherjee}, together
with the well known form of the first term in the Seiberg-Witten expansion of
the classical action for an abelian gauge field, speaks in favor of the
generality of the action~\eqref{action}.

\section{\label{sec:fieldeqs}Field equations and conformal ansatz}

Varying the action~\eqref{action} with respect to the metric and the gauge
field we obtain the field equations. Substituting in them the expansions
\begin{equation}
  g_{\a\b} = g^{(0)}_{\a\b} +  g^{(1)}_{\a\b} + \ldots 
  \qquad
  F_{\a\b} = F^{(0)}_{\a\b} +  F^{(1)}_{\a\b} + \ldots 
\label{expansions}
\end{equation}
for $g_{\a\b}$ and $F_{\a\b}$ in powers of $\th^{\a\b}$, retaining
contributions up to order one and identifying coefficients of the same
order, we obtain
\begin{eqnarray}
   & \bR_{\a\b} = \ka^2 \bT_{\a\b}  &  \label{E0} \\[3pt]
   & \bna_\a\, \bF^{\a\b} = 0 & \label{F0}   
\end{eqnarray}
for the Einstein and field equations at order one, and
\begin{eqnarray}
   & \hR_{\a\b} =  \ka^2 \hT_{\a\b}
   & \label{E1} \\[3pt]
   &\bna_\a \big( \hF^{\a\b} + \hH^{\a\b}\big) = L^\b & \label{F1}  
\end{eqnarray}
for the Einstein and field equations at order one.  Here the notation is as
follows. Quantities of order zero in $\th^{\a\b}$ are denoted with a bar and
quantities of order one with a hat, so that \hbox{$\bg_{\a\b}\!=\!
  g^{(0)}_{\a\b}\,$}, \hbox{$\hg_{\a\b}\!=\! g^{(1)}_{\a\b}\,$}, etc. The
order-zero contribution to the Ricci tensor is constructed from the metric
$\bg_{\a\b}$, while the order-zero contribution to the energy-momentum tensor
reads
\begin{equation*}
   \bT_{\a\b} = \bF_\a^{~\m} \bF_{\b\m} 
        - \frac{1}{4}\, \bg_{\a\b}\,\bF_{\m\n} \bF^{\m\n}\,.
\end{equation*}
In turn, the order-one contributions $\hR_{\a\b}$ and $\hT_{\a\b}$ to
the Ricci and the energy-momentum tensors have the form
\begin{align}
  &\qquad\qquad\qquad\quad \hR_{\a\b} = \frac{1}{2} \> \bg^{\ga\dl} \,
    \big( \bna_{\!\a} \bna_{\!\b}\, \hg_{\ga\dl} 
        - \bna_{\!\ga} \bna_{\!\b}\, \hg_{\a\dl}
        - \bna_{\!\ga} \bna_{\!\a}\, \hg_{\b\dl}  
        + \bna_{\!\ga} \bna_{\!\dl}\, \hg_{\a\b} \big)  \nonumber \\
   & \hT_{\a\b} = \Big[  \hK_{(\a}^{~~\m} + \hF_{(\a}^{~~\m} - 
        \frac{1}{2}\,\hg_\rho^{~\m} \bF_{(\a}^{~~\rho}\, \Big]\, 
            \bF^{\vphantom{\m}}_{\b)\m} 
      - \frac{1}{2}\,\bg_{\a\b}\, \big( \hK_{\m\n} + \hF_{\m\n}
      - \bF_{\m}^{~\rho}\, \hg_{\rho\n} \big)\, \bF^{\m\n}
      - \frac{1}{4}\,\hg_{\a\b}\, \bF_{\m\n} \bF^{\m\n}\,, \nonumber 
\end{align}
where we have defined 
\begin{equation*}
  \hK_{\a\b} = \th^{\m\n} \Big( \bF_{\m\a} \bF_{\n\b} 
        -\frac{1}{4}\, \bF_{\m\n}  \bF_{\a\b} \Big) \,,
\end{equation*}
and introduced the notation $X_{(\a}Y_{\b)}=X_\a Y_\b + X_\b Y_\a$.  In
eq.~\eqref{F1} $\hH^{\a\b}$ and $\hat{L}^\b$ stand for
\begin{eqnarray*} 
   & \hH^{\a\b} =  \th^{\m\n} \big( \bF_\m^{~\a} \bF_\n^{~\b} 
     - {\ds \frac{1}{2}}\, \bF_{\m\n} \bF^{\a\b}\big) 
    - \big( \th^{\a\m} \bF^{\b\n} - \th^{\b\m} \bF^{\a\n} \big)
      \bF_{\m\n} 
    - {\ds \frac{1}{4}}\, \th^{\a\b} \bF_{\m\n} \bF^{\m\n} & \\[3pt] 
   & \hat{L}^\b = \bg^{\m\b} \bna_\a \big( \hg_{\m\n} \bF^{\a\n}\big)
      - \bg^{\m\a} \bna_\a \big( \hg_{\m\n} \bF^{\b\n} \big)
      - \hat{\gm}^{\a}_{\a\m} \bF^{\m\b}\,, &
\end{eqnarray*}
with the order-one contribution $\hat{\gm}^\a_{\b\ga}$ to the Christoffel
symbol given by
\begin{equation}
   \hat{\gm}^\a_{\b\ga} = {\ds \frac{1}{2}} \, \bg^{\a\dl}\, 
      \big( \bna_\b \,\hg_{\dl\ga} + \bna_\ga \,\hg_{\b\dl} 
          - \bna_\dl\, \hg_{\b\ga}\big) \,. 
\label{1-Chris}
\end{equation}
In these expressions indices are lowered, raised and contracted with the
zeroth order metric $\bg_{\a\b}$, so that
\hbox{$\,\bF^{\a\b}\!=\bg^{\a\m}\,\bg^{\b\n}\bF_{\m\n}\,$},
\hbox{$\,\hT^\a_{~\a}\!=\bg^{\a\b}\hT_{\a\b}\,$}, etc.
  
It may look at first sight that expanding the field strength $F_{\a\b}$ in
eq.\eqref{expansions} in a power series of $\th^{\a\b}$ clashes with the idea
that the first order noncommutative correction to $\bF_{\a\b}$ is determined
by the Seiberg-Witten map. Note, however, that $\hT_{\a\b}$ will in general
produce an order-one correction to the metric, which in turn may react back
and modify the field strength. Solutions with $\hF_{\a\b}=0$ will correspond
to the case in which the Seiberg-Witten order-one noncommutative correction
remains unchanged by gravity and all noncommutative corrections to
$\bF_{\a\b}$ are provided by the map itself.

In eqs.~\eqref{E0} and \eqref{F0} we recognize the field equations for the
coupling of a conventional Abelian gauge field to gravity. This problem has
been extensively studied in the literature, see e.g.~\cite{Stephani}. The
order-one noncommutative corrections $\hg_{\a\b}$ and $\hF_{\a\b}$ are in turn
governed by eqs.~\eqref{E1} and \eqref{F1}. We will look for solutions for
$\hg_{\a\b}$ of the form
\begin{equation}
  \hg_{\a\b} = \ho\,\bg_{\a\b}\,,
\label{ansatz}
\end{equation}
with $\ho$ a function of $x^\a$ of order one in $\th^{\a\b}$. For this ansatz,
the first order contribution to the energy-momentum tensor takes the form
\begin{equation}
   \hT_{\a\b} =  \Big[ \hK_{(\a}^{~~\m} + \hF_{(\a}^{~~\m} - 
        \frac{\ho}{2}\,\bF_{(\a}^{~~\m}\, \Big]\, \bF^{\phantom{\m}}_{\b)\m} 
    -  \frac{1}{2}\, \bg_{\a\b}\, \big( \hK_{\m\n} + \hF_{\m\n} 
               - {\ds \frac{\ho}{2}}\,\bF_{\m\n}\big) \bF^{\m\n}\,, 
\label{T-conformal}
\end{equation}
while, very importantly, $\hat{L}^\b$ vanishes. The first order
equations~\eqref{E1} and \eqref{F1} then read
\begin{eqnarray}
   & \bna_{\!\a} \bna_{\!\b}\, \ho= \ka^2 \hT_{\a\b} & \label{E-conformal}\\
   & \bna_{\!\a} \,\big( \hF^{\a\b} + \hH^{\a\b}\big) = 0 
   & \label{F-conformal}
\end{eqnarray}
The function $\ho$ satisfies the constraint~$\bna^2\ho=0$. This is so since
the order-one contribution to the Ricci tensor is $\hR_{\a\b}\!=\!
\bna_{\!\a} \bna_{\!\b}\, \ho +\frac{1}{2}\bg_{\a\b}\bna^2\ho$ and
$\hT_{\a\b}$ is traceless, so that is $\hR^\a_{~\a}$ vanishes. The way to
proceed is now clear. Take a solution of the order-zero equations~\eqref{E0}
and~\eqref{F0}, substitute it in eqs.~\eqref{E-conformal}
and~\eqref{F-conformal}, and solve the resulting equations for $\ho$ and
$\hF_{\a\b}$. In the next sections we will find solutions to these equations
describing noncommutative modifications of {\it pp\,}-waves.

\section{\label{sec:general}Noncommutative  \textit{pp\,}-waves. General
  considerations} 

As is known~\cite{Stephani}, all homogeneous null Einstein-Maxwell fields are
represented by the {\it pp\,}-wave spacetime metric
\begin{equation}
  ds^2 =   2\,du\,dv + 2\,H(u,x,y)\, du^2 - dx^2 -dy^2 \,,
\label{metric}
\end{equation}
where the coefficient $H$ has one of the following two forms
\begin{align}
    {\rm case~0\!:}~~&
   H=\frac{b^2\si^2}{2}\,(x^2+y^2) + a \si^2
      \big[ (x^2-y^2)\,\cos(2 c \si u) + 2xy\,\sin(2c  \si u) \big]
   \label{case-0} \\[3pt]
    {\rm case~1\!:}~~&
   H=\frac{b^2}{2\,u^2}~(x^2+y^2) + \frac{a}{u^2}~
      \big[ (x^2-y^2)\,\cos\big(2 c \ln(\si u)\big)
          + 2xy\,\sin\big(2c \ln(\si u)\big) \big]\,.
   \label{case-1}
\end{align}
The EM field, in turn, has as only nonzero components
\begin{alignat}{3}
   & {\rm case~0}:&~~
   & \bF_{ux} = \bE = \frac{\sqrt{2}\,b\,\si}{\ka}\, \cos\phi(\si u)
   &
   & \bF_{uy} = \bB = -\,\frac{\sqrt{2}\,b\,\si}{\ka}\,\sin\phi(\si u)
   \label{case-0-EM} \\[3pt]
   & {\rm case~1}:&~~  & \bF_{ux} = \bE
      = \frac{\sqrt{2}\,b}{\ka u}\, \cos\phi\big(\ln(\si u)\big)
   & \qquad & \bF_{uy} = \bB
     = -\,\frac{\sqrt{2}\,b}{\ka u}\,\sin \phi\big(\ln(\si u)\big)\,.
   \label{case-1-EM}
\end{alignat}
In these equations, \hbox{$u\!=\!t-z$}, \hbox{$\,v\!=\!t+z$}, $\,x$ and $\,y$
are spacetime coordinates, $\si$ is an arbitrary mass scale, $a,b,c$ are
arbitrary real constants, and $\phi$ is an arbitrary function of its argument.
In case~0, the coordinate $u$ may take on arbitrary values, whereas in case~1
one has $u\!>\!0$. The only nonvanishing Christoffel symbols are
\begin{equation*}
   \bar{\gm}^v_{uu}=\pa_u H \qquad  \bar{\gm}^v_{ui}=\pa_i H 
   \qquad \bar{\gm}^i_{uu}=\pa_i H \qquad (i=x,y)\,.
\end{equation*}
It is worth noting that the change $\si u^\prime\!=\!\ln(\si u)$ in case~1
yields for $\bF_{\a\b}$ and $H$ the same expressions as in case~0, but changes
the metric coefficient $\bg_{uv}$ from 1 to $e^{\si u^\prime}$. Here we will
not make this change and stick to the notation presented above. The reason for
this it that it is customary in the literature to write {\it pp\,}-waves in
coordinates systems for which the $uv$-metric coefficient is 1. We finally
note that the vector $\,\bar{\xi}_\a\!=\!\dl_{\a u}\,$ is null
$\,(\bar{\xi}^2\!=\!0)\,$ and covariantly constant
$\,(\bna_{\!\a}\,\bar{\xi}_\b\!=\!0)\,$ and that the the null surfaces of the
metric are $u={\it const}$. As regards the EM field $\bF_{\a\b}$, it is
obvious that it is null $\bF^{\a\b}\bF_{\a\b}=0$ and that it needs not to be a
plane wave, since the function $\phi$ in \eqref{case-0-EM} and
\eqref{case-1-EM} is arbitrary. Plane EM waves correspond to $\,\phi(w)= w$.

We take $\bg_{\a\b}$ and $\bF_{\a\b}$ as above in the reference frame in which
$\th^{\a\b}$ has constant components. It is straightforward to show that
\begin{equation}
   \th^{\m\n}\bF_{\m\a}\,\bF_{\n\b} = \frac{1}{2}\> 
       \th^{\m\n}\bF_{\m\n}\,\bF_{\a\b}\,,
\label{curious}
\end{equation}
which in turn implies
\begin{equation*}
   \hK_{\a\b} = \frac{1}{4}\> \th^{\m\n}\bF_{\m\n}\,\bF_{\a\b}
   \qquad
   \hH_{\a\b} =  (\th_{\a}^{~\m}\bF^\n_{~\b} 
      - \th_\b^{~\m}\bF^\n_{~\a})\,  \bF_{\m\n}\,.
\end{equation*}
Furthermore, using the notation
$\,\th\hspace{-1.5pt}\cdot\hspace{-2.2pt}L=\th^{\a\b}L_{\a\b}\,$ for any
tensor $L_{\a\b}$, and defining
\begin{equation*}
   e_x=\th^{ux} \qquad e_y=\th^{uy}\,,
\end{equation*}
so that $\,\thbF\!=\! 2\,(e_x\bE+e_y\bB)$, we have that the only nonvanishing
components of $\hK_{\a\b}$ and $\hH_{\a\b}$ are
\begin{alignat*}{2}
    & \hK_{ux} = \frac{1}{4}\> (\thbF)\,\bE  &\qquad &
      \hK_{uy} = \frac{1}{4}\> (\thbF)\,\bB \\[3pt]
    & \hH_{ux} = (\bE^2+\bB^2)\, e_x & \qquad & 
      \hH_{uy} = (\bE^2+\bB^2)\,e_y \,.
\end{alignat*}

We will look for solutions $\ho$ that only depend on $u$. The metric,
including first order noncommutative corrections, is given by 
\begin{equation}
   g_{\a\b}=(1+\ho)\,\bg_{\a\b} \,.
\label{full-metric}
\end{equation}
This yields the following order-one nonzero contributions
$\hat{\gm}^\a_{\b\ga}$ to the Christoffel symbols:
\begin{equation*}
  \hat{\gm}^u_{uu} = \pa_u\ho\qquad 
  \hat{\gm}^v_{uu} = -H\, \pa_u\ho\qquad 
  \hat{\gm}^v_{uv} = \frac{1}{2}\>\pa_u\ho \qquad 
  \hat{\gm}^i_{uj} = \frac{1}{2}\>\dl^i_{\,j}\,\pa_u\ho \,.
\end{equation*}
Let us see that the metric~\eqref{full-metric} is a {\it pp\,}-wave.
Since, by definition~\cite{Stephani}, a spacetime metric is a {\it pp\,}-wave
if it admits a null and covariantly constant vector, we must find one such
vector for the metric~\eqref{full-metric}. It is clear that
$\,\xi_\a\!=(1+\ho)\,\dl_{\a u}\,$ does the job. Indeed, nullity $\xi^2\!=\!0$
is obvious. As regards covariant constancy, recalling that 
$\,\bar{\xi}_\a\!=\!\dl_{\a u}$ was covariantly constant with respect to
$\bg_{\a\b}$ and noting the expressions for $\bar{\gm}^\a_{\b\ga}$ and
$\hat{\gm}^\a_{\b\ga}$, we have up to order one
\begin{equation*}
   \na_{\!\a}\xi_\b = \bna_{\!\a} \,\bar{\xi}_\b 
    +  \pa_\a\,\big( \ho\,\dl_{\b u}\big) 
    -  \bar{\gm}^\ga_{\a\b}\,\ho\,\dl_{\ga u} 
    -  \hat{\gm}^\ga_{\a\b}\,\dl_{\ga u} = 0\,.
\end{equation*}
To write the metric~\eqref{full-metric} in standard {\it pp\,}-wave
coordinates, we perform the change $\,u\to \tilde{u}$, with $\tilde{u}$ defined
by the differential equation
\begin{equation}
   (1+\ho)\, du = d\tilde{u}\,.
\label{tilde-u}
\end{equation}
The metric then reads
\begin{equation*}
  ds^2 = 2\,d\tilde{u} dv 
       + \frac{2}{1+\ho(\tilde{u})}~H(\tilde{u},x,y)\,d\tilde{u}^2 
       - \big[1+\ho(\tilde{u})\big]\,(dx^2 +dy^2)
\end{equation*}
and has $\,\tilde{u}\!=\!{\it const}$ as null surfaces. In this new coordinate
system, noncommutativity does not explicitly enter in the characterization of the
null spacetime surfaces. However, the separation between points in different
null surfaces ($d\tilde{u}\!\neq\!0$) and between points within the same null
surface ($dx,dy\!\neq\!0$) does in general depend on the noncommutative
parameters $\th^{\a\b}\!$. We will see below explicit realizations of this.
Under the coordinate change~\eqref{tilde-u}, $\hF_{\a\b}$ and $\th^{\a\b}$
transform as usual tensors, so that for example
$\,\tilde{\th}^{ux}\!=\![1+\ho(\tilde{u})]\,\th^{ux}$, and $\,\thbF\,$ and
$\,\thbf\,$ remain invariant.

The explicit form of $\ho$ is to be found from the Einstein equations
\eqref{E-conformal}, which now take the form
\begin{alignat}{2}
   \frac{d^2\ho}{du^2} & =  \ka^2 \hT_{\a\b}  & \qquad
     & {\rm if}~\a=\b=u   \label{E-conformal-uu}\\[3pt]
    0 & = \hT_{\a\b} & & {\rm otherwise}\,. \label{E-conformal-other}
\end{alignat}
As concerns the noncommutative corrections $\hF_{\a\b}$ to the EM field, it is
straightforward to see that the expression~\eqref{T-conformal} for
$\hT_{\a\b}$ and eq.~\eqref{E-conformal-other} imply that, except for
$\hF_{ux}$ and $\hF_{uy}$, all other components of $\hF_{\a\b}$ vanish.  Hence
$F_{\a\b}\!=\!\bF_{\a\b}\!+\!\hF_{\a\b}\,$ has $ux$ and $uy$ as only nonzero
components, and is null for~\eqref{full-metric}.

\section{\label{sec:explicit}Noncommutative  \textit{pp\,}-waves. Explicit
  solutions} 

In this section we consider some solutions for $\hF_{\a\b}$ of physical
interest. We will see that, while the solution for $\ho$ is unique, there is
a large arbitrariness in the solution for $\hF_{\a\b}$.

\subsection{\label{sec:vacuum}Vacuum noncommutative metric corrections}

Let us first consider $\hT_{\a\b}=0$, corresponding to vacuum noncommutative
metric corrections. For $\hT_{\a\b}=0$, we demand
\begin{equation}
  \hF_{\a\b} =\,  \frac{\ho}{2} \> \bF_{\a\b} - \hK_{\a\b}\,.
\label{F-vacuum-first}
\end{equation}
Now, $2\hK_{\a\b}$ and $\hH_{\a\b}$ are not energetically distinguishable
from each other, since they both give the same contribution to the
energy-momentum tensor. We therefore replace $2\hK_{\a\b}$ in
eq.~\eqref{F-vacuum-first} with a linear combination of $2\hK_{\a\b}$ and
$\hH_{\a\b}$ that gives the same contribution to $\hT_{\a\b}$ as
$2\hK_{\a\b}$. That is,
\begin{equation}
 \hF_{\a\b} =  \frac{\ho}{2} \> \bF_{\a\b}  
      - \frac{1}{2}\> \big[ 2a_1\hK_{\a\b} + (1-a_1)\,\hH_{\a\b}\big] \,.
\label{F-vacuum-second}
\end{equation}
Here $a_1$ may even be regarded, not as a constant, but as a function of $u$,
for it does not enter $\hT_{\a\b}=0$. Following this line of argumentation,
one could think of including in $\hF_{\a\b}$ other antisymmetric two-tensors
with the same contribution to the energy-momentum tensor, or with different
contributions but such that they cancel among themselves. Using
eq~\eqref{curious} and
\begin{equation*}
  \bF_{\a\m} \,\bF^\m_{~\,\b}=(\bE^2+\bB^2)\>\dl_{\a u}\,\dl_{\b u} \qquad 
  \bF_{\a\m}\,\bF^{\m\n}\bF_{\n\b}=0  \qquad
  \th^{\m\n} (\bF_{\m\a}\,\bF_{\b\rho} - 
         \bF_{\m\b}\,\bF_{\a\rho}) \,\bF_\n^{~\rho}=0\,,
\end{equation*}
it is not difficult to see, however, that the only nonzero antisymmetric
two-tensors that can be formed with one $\th^{\a\b}$ and two or more
$\bF_{\a\b}$ are $\hK_{\a\b}$ and $\hH_{\a\b}$. Hence we stop
at~\eqref{F-vacuum-second}, which can be recast as
\begin{equation}
   \hF_{\a\b} =  \frac{1}{2}\> \big( \ho 
                     - \frac{a_1}{2}~ \thbF\big)\, \bF_{\a\b}  
      - \frac{1}{2}\> (1-a_1)\,\hH_{\a\b}\,.
\label{F-vacuum}
\end{equation}

{}For this ansatz, the field equation~\eqref{F-conformal} is trivially
satisfied for all $a_1$ and the Einstein equation~\eqref{E-conformal-uu} takes
the \hbox{form~$\,\ho^{\,\prime\prime}\!=\!0$}, where the prime denotes
differentiation with respect to $u$.
In what follows we show that its solution is
\begin{equation}
   \ho = (\thbf)\, (k_1 + k_2 \,\si u)\, ,
\label{omega-vacuum}
\end{equation}
where $k_1$ and $k_2$ are arbitrary real constants and $\thbf$ is the
contraction of $\th^{\a\b}$ with an antisymmetric tensor $\bar{f}_{\a\b}$
whose only nontrivial components are
\begin{equation}
  \bar{f}_{ux} = \frac{b\si}{\ka}\>\cos\eta \qquad   
  \bar{f}_{uy} = \frac{b\si}{\ka}\> \sin\eta \,,
\label{constant-f}
\end{equation}
with $\eta$ a real arbitrary constant.  To prove eq.\eqref{omega-vacuum}, we
proceed in three steps:

{\leftskip=24pt
\noindent
\hskip -11pt $\bullet$ {\sl Step 1.} We first note that the solution
to~\hbox{$\,\ho^{\,\prime\prime}\!=\!0$} is $\,\ho= \hk_1 + \hk_2 u$, with
$\hk_1$ and $\hk_2$ integration constants.  Since $\hk_1$ and $\hk_2$ are of
order one in $\th^{\a\b}$, they must be contractions of $\th^{\a\b}$ with
antisymmetric tensors of mass dimension 2 taking constant values in the
reference frame that we are considering
  
\noindent
\hskip -11pt $\bullet$ {\sl Step 2.} Next we show that all tensors of that
type for the background metric~\eqref{metric} have the
form~\eqref{constant-f}.  To this end, we recall~\cite{Ehlers} that
covariantly constant null antisymmetric two-tensors exist if and only if
space-time is a {\it pp\,}-wave. In addition, metrics that cannot be
decomposed into the product of two two-dimensional metrics do not admit
covariantly constant non-null antisymmetric two-tensors~\cite{Debever}. Since
these two conditions concur for the metric~\eqref{metric}, all covariantly
constant antisymmetric two-tensors in our case are null. Let us take one such
tensor, which we denote by $\bar{f}_{\a\b}$, and demand it to be constant, so
that $\pa_\ga\bar{f}_{\a\b}\!=\!0$. Covariant constancy
$\bna{\!\ga}\bar{f}_{\a\b}\!=\!0$ then reduces to
\begin{equation*}
    \bar{\gm}^\m_{\ga\a} \bar{f}_{\m\b} 
     + \bar{\gm}^\m_{\ga\b} \bar{f}_{\a\m} = 0\,.
\end{equation*}
We remind ourselves at this point that a constant and at the same time
covariantly constant null antisymmetric two-tensor $\bar{f}_{\a\b}$ can be
written as~\cite{Stephani}
\begin{equation*}
  \bar{f}_{\a\b} = \bar{q}_\a \bar{p}_\b - \bar{p}_\a \bar{q}_\b \,
\end{equation*}
with $\bar{q}_\a$ and $\bar{p}_\a$ null and spacelike constant vectors
satisfying
\begin{equation*}
  \bar{q}^2 = 0 \qquad \bar{\gm}^\ga_{\a\b}\,\bar{q}_\ga =0 
  \qquad \bar{p}^2=-1  \qquad \bar{q}\cdot\bar{p}=0\,.
\end{equation*}
It is now simple algebra to prove that the only solution to these equations is
$\,\bar{q}_\a\!=\!(q,0,0,0)$ and
\hbox{$\,\bar{p}_\a=\!(0,0,\cos\eta,\sin\eta)$}, with $q$ and $\eta$ arbitrary
real numbers. 

\noindent
\hskip -11pt $\bullet$ {\sl Step 3.} We observe that the constants $\hk_1$ and
$\hk_2$ must be zero if the EM background $\bF_{\a\b}$ vanishes, for then one
is left with the usual Einstein equations in vacuum, in which no $\th^{\a\b}$
is involved. Based on this and the observation that $\bF_{\a\b}$ has the form
of $\bar{f}_{\a\b}$ constructed in step 2, we take $q=b\si/\ka$ and
obtain~\eqref{constant-f}.
\par}

Let us now go back to $\ho$ in eq.~\eqref{omega-vacuum}. Performing the change
$\,u\to \tilde{u}$ in eq.~\eqref{tilde-u}, we obtain
\begin{equation*}
    1 +\ho =\sqrt{\big[1+ k_1\>(\thbf)\big]^2 + 2\,k_2\>(\thbf) 
                                                \,\si\tilde{u}}
\end{equation*}
For $k_2\!=\!0$ noncommutativity is felt in the same way in all null surfaces
$\tilde{u}\!=\!{\it const}$. On the other hand, for $k_2\!\neq\!0$, null
surfaces sort of dilate, since up to order one, the distance between two
points on the same null surface $\,\tilde{u}\!=\!\tilde{u_0}\,$ is
\begin{equation*}
  ds^2_{\rm null} = \big[1+(\thbf)\,(k_1 +k_2\,\si\tilde{u}_0)\big]\, 
         (dx^2+dy^2)\,.
\end{equation*}
We postpone to Section VI a detailed discussion of the noncommutative
correction $\hF_{\a\b}$ to the EM background field given by
eq.~\eqref{F-vacuum}.

\subsection{\label{sec:nonvacuum}Nonvacuum noncommutative metric corrections}

We next consider for $\hF_{\a\b}$ the ansatz
\begin{equation}
    \hF_{\a\b} =  \frac{1}{2}\,(\ho -\hphi)\, \bF_{\a\b} 
    - \frac{1}{2}\,(1-a_1) \hH_{\a\b} \,,
\label{general}
\end{equation}
where $\hphi$ is a function of $u$ or order one in $\th^{\a\b}$ to be
determined.  This $\hF_{\a\b}$ is a non-vacuum generalization of that
considered in eq.~\eqref{F-vacuum}.  With respect to the latter, we have
replaced the prefactor $\,\thbF\,$ in front of $\bF_{\a\b}$ with a function
$\hphi$.  The motivation for doing this is that $\ho$ may contribute, through
the Einstein equation~\eqref{E-conformal-uu}, to $\hF_{\a\b}$ with a
term proportional to $\bF_{\a\b}$ whose prefactor may not of the form
$\thbF$, and the idea is that $\hphi$ accounts for such a contribution. It is
straightforward to see that the field equation~\eqref{F-conformal} holds
trivially for all $\hphi$ and $a_1$. By suitably choosing $\hphi$ and $a_2$,
we may construct a large variety of solutions. Let us consider some of them.

{\sl Example 1}. Assume $\hphi\!=\!0$. If $a\,_1\!=\!0\,$ we fall into the
case studied in the previous subsection, so we will consider here $a_1\!\neq\!
0$.  The Einstein equation~\eqref{E-conformal-uu} then reduces to
\begin{equation*}
   \frac{d^2\ho}{du^2} = -\,a_1 \ka^2~(\bE^2+\bB^2)\>
        (\thbF)\,.
\end{equation*}
Its solution is given by
\begin{equation*}
   \ho =  (\thbf)\, (k_1 + k_2\,\si u) +\ho_{\rm p}\,,
\end{equation*}
with $\ho_{\rm p}$ a particular solution of the complete equation. Different
choices for $\phi$ in eqs.~\eqref{case-0-EM},~\eqref{case-1-EM} will yield
different particular solutions. Of special relevance are plane EM waves with
wave fronts $u\!=\!{\it const}$, for which $\phi(w)\!=\! w$.  In this
case, it is straightforward to see then that $\ho_{\rm p}$ is given by
\begin{equation*}
   {\rm Plane~EM~wave\!:}~~ \phi(w)\! = \! w ~~\left\{
\begin{array}{ll}
    {\rm case~0\!:}& ~ {\ds \ho_{\rm p}=  2a_1b^2 ~(\thbF)} \\[3pt]
    {\rm case~1\!:}& ~ {\ds \ho_{\rm p}= -\,\frac{a_1b^2}{5}
                          ~(\thbF + 3~ \thdbF)\,,}
\end{array}
                                               \right. 
\end{equation*}
where $\,\thdbF\,$ is the contraction of $\th^{\a\b}$ with the Hodge dual of
$\bF_{\a\b}$. If we now substitute the solution for $\ho$ in the
expression~\eqref{general} for $\hF_{\a\b}$, we again obtain a sum two dipolar
polarization contributions (one linear and one nonlinear) and an
induction/displacement contribution.

{\sl Example 2}. Take now $\,\hphi\!=\!\ho+(\tfrac{1}{2}\,a_1 + a_2)\>\thbF\,$,
with $a_2$ an arbitrary real constant. The Einstein
equation~\eqref{E-conformal-uu} then takes the form
\begin{equation*}
   \frac{d^2\ho}{du^2} - \ka^2\, (\bE^2+\bB^2)\,\ho 
     = a_2\,\ka^2 \, (\bE^2+\bB^2)\>(\thbF)\,.
\end{equation*}
This is again a second order differential equation that can be solved without
difficulty. For a  plane wave EM background, the solution is given by
\begin{equation*}
\begin{array}{ll}
  {\rm case~0\!:}& ~ {\ds \ho =  
           (\thbf)\,\big( k_1\, e^{\sqrt{2}\,b\si u} 
         + k_2\, e^{-\sqrt{2}\,b\si u}\, \big) 
         -\frac{2\,a_2 b^2}{\,(2b^2+1)}~\thbF }\\[9pt]
  {\rm case~1\!:}& ~ {\ds \ho = 
           (\thbf)\,\big[ k_1\,{(\si u)}^{\la_+} 
         + k_2\, {(\si u)}^{\la_-}  \big] 
         - \frac{a_2\,b^2}{5-2b^2+2b^4}~
         \big[ (2b^2-1)\>\thbF - 3\>\thdbF\big] \,,}
\end{array}
\end{equation*}
with $\,2\la_\pm=1\pm\sqrt{1+8b^2}$. In case 0, the range of variation for the
coordinate $u$ goes from $-\infty$ to $+\infty$. The homogeneous part of $\ho$
then dominates and as $u\to\pm\infty$ the null surfaces inflate. This is in a
way a runaway configuration, since $\ho$ ends up growing indefinitely. Similar
considerations apply to case 1 for $u\to 0,\infty$. The main difference with
the solutions for $\hF_{\a\b}$ studied in previous sections is that now there
is not a linearly polarized contribution.  Note, as a particular case, that if
$a_1\!=\!1$ and $a_2\!=\!-\tfrac{1}{2}$, $\hF_{\a\b}$ in~\eqref{general}
vanishes, so that the gravitational field does not react back on the EM
background.

\section{\label{constitutive} Constitutive relations and comparison with
  noncommutative electromagnetic  plane  waves in flat spacetime}

The complete EM field, which we denote by $F_{\a\b}$, is the sum of the
order-zero $\bF_{\a\b}$ and order-one $\hF_{\a\b}$ contributions. For the sake
of concreteness, we will consider the case discussed in Subsection V~A and
furthermore we will take \eqref{case-0} and \eqref{case-0-EM} as background.
Similar considerations apply to other cases.  Putting together the background
EM contributions and the noncommutative corrections in eq.~\eqref{F-vacuum},
we obtain
\begin{align}
   D = F_{ux} & = \bE +  \frac{\ho}{2}\>\bE 
      - \frac{a_1}{4}\>(\thbF)\>\bE
      - (1-a_1)\>\frac{b^2\si^2}{\ka^2}~e_x \label{const-D}\\[3pt]
   H = F_{uy} & = \bB +  \frac{\ho}{2}\>\bB 
      - \frac{a_1}{4}\>(\thbF)\>\bB
      - (1-a_1)\>\frac{b^2\si^2}{\ka^2}~e_y \label{const-H}
\end{align}
as the only nonvanishing components of the complete EM field. These define
the displacement $\bold{D}$ and induction $\bold{H}$ vectors in the same way
that the components of $\bF_{\a\b}$ define the electric $\boldE$ and magnetic
$\boldB$ background fields. Indeed, in a coordinate system $(t,x,y,z)$,
$\boldE$ and $\boldB$ have Cartesian components $\,E_i=\!\bF_{0i}\,$ and
$\,B_i\!=\!-\bF_{jk}$ with $ijk$ a cyclic permutation of $123$.  Similarly,
for the components of $\bold{D}$ and $\bold{H}$ we have $\,D_i=\!F_{0i}\,$ and
$\,H_i\!=\!-F_{jk}$.  Eqs.~\eqref{const-D},~\eqref{const-H} are thus
constitutive relations and can be easily inverted. Let us analyze the terms
occurring in them.

The first term in eqs.~\eqref{const-D},~\eqref{const-H} is the EM background
$\bF_{\a\b}$ contribution. The second one arises from $\ho\bF_{\a\b}$ in
eq.~\eqref{F-vacuum} and is a susceptibility contribution due to gravity, with
susceptibility coefficient $\ho$. This contribution is inhomogeneous ($\ho$ is
a function of $u$) and depends on the properties of the gravitational field
($\ho$ is proportional to $\,\thbf$, with $\bar{f}_{\a\b}$ determined by the
{\it pp\,}-wave background geometry). The third term
in~\eqref{const-D} and \eqref{const-H} comes from $(\thbF)\bf_{\a\b}$
in~\eqref{F-vacuum}, is quadratic in the background EM field and has its 
origin in the Seiberg-Witten map, since this was used to construct the
classical action and hence enters the Einstein and field equations. Even more,
the order-one contribution to the field strength provided by the
Seiberg-Witten map
\begin{equation*}
  \th^{\m\n} \big[ \bF_{\m\a} \bF_{\n\b} -
       (\pa_\m\bF_{\a\b})\bar{A}_\n \big]
\end{equation*}
can be recast for our {\it pp\,}-wave background geometry as
\begin{equation*}
  (\thbF)  \bF_{\a\b} - \frac{1}{\sqrt{-g}}~
    \bna_{\!\m}\,(\th^{\m\n} \bA_\n \bF_{\a\b})\,,
\end{equation*}
which, modulo the total derivative in the second term, is the nonlinear
contribution that we are discussing. Finally, the last contribution
in~\eqref{const-D} and \eqref{const-H} can be regarded as a constant
polarization/induction that can be tuned at will by adjusting $a_1$. This
contribution is dynamically generated and arises because, as already mentioned
in Subsection V~B, $2\hK_{\a\b}$ and $\hH_{\a\b}$ cannot be energetically
distinguished from each other and both satisfy the order-one field
equation~\eqref{F-conformal}.  Even if one takes
\hbox{$\,\phi\!=\!0~(\tfrac{\pi}{2})$}, so that the magnetic (electric)
component $\bB~(\bE)$ of the EM background is zero, the complete EM field
acquires through this term an induction (displacement) component proportional
to $e_y~(e_x)$.

The nonlinear contribution in eqs.~\eqref{const-D},~\eqref{const-H} is
responsible for harmonic generation similar to that in nonlinear optics. To
understand this, we consider this contribution together with the last one
in~\eqref{const-D},~\eqref{const-H}, and take $\phi(\si u)=\si u$ in the
background EM field. Doing so, we obtain
\begin{equation*}
{\rm 3rd + 4th~contributions} \,=
     \big(\>\frac{a_1}{2}-1\big)\>\frac{b^2\si^2}{\ka^2} ~ 
         \begin{bmatrix} \,e_x\, \\ e_y \end{bmatrix}
     + \frac{a_1}{2}\>\frac{b^2\si^2}{\ka^2} ~
         \begin{bmatrix} \>e_y\sin(2\si u) - e_x\cos(2\si u)\> \\
           e_x\sin(2\si u) + e_y\cos(2\si u)\end{bmatrix} \,.
\end{equation*}
The first term on the right-hand side is again a constant
polarization/induction contribution, whereas the second one describes an EM
plane wave propagating in the same direction as the background wave but with
twice its frequency. Hence harmonic generation and constant
polarization/induction cannot be distinguished energetically from each other
and one can ``move'' between them by tuning $a_1$.

It is important to remark that our solutions for the complete EM field, though
similar, are of a different type to those discussed in the literature for flat
spacetime~\cite{Guralnik,Abe}. The difference goes beyond the explicit form of
the solutions and affects the nature of the EM waves.  The susceptibility
contribution due to gravity that we have found is absent in the flat spacetime
solutions. The nonlinear contribution, having its origin in the Seiberg-Witten
map, is present in both cases. Finally, whereas in our solution there is a
dynamically generated constant polarization/induction at order one in
$\th^{\a\b}$, in the flat spacetime solutions presented in~\cite{Guralnik,Abe}
there is not such a contribution.  One must bear in mind, though, that the
argument above for harmonic generation shows that the nonlinear contribution
can be split into a constant piece and a higher harmonic term. The difference
is that whereas in our case the coefficient of the constant contribution is
$(a_1-2)/2$, with $a_1$ is arbitrary, in flat spacetime it takes the value
$-1/2$.

By contrast, the flat spacetime solution for the EM field has an order-zero
constant contribution. This generates through the non linear term a
longitudinal component for the the electric and magnetic fields which
propagates at different velocity than the transverse
components~\cite{Guralnik,Cai}. Nothing of this happens in our case, since such an
order-zero contribution cannot occur in our background solution. The reason
for this is that if a constant contribution is added to a given background
solution $\bF_{\a\b}$ the Einstein equation does not hold and such a
background is not acceptable.

\section{\label{sec:outlook}Conclusion and outlook}

In this paper we have considered the noncommutative coupling of gravity to an
EM field as described by Seiberg-Witten type formal series in the
noncommutativity parameters $\th^{\a\b}$. We have constructed a very large
class of null {\it pp\,}-waves that solve the corresponding field equations up
to order one in $\th^{\a\b}$. These solutions can thus be regarded as the
first-order noncommutative contributions to full noncommutative
Einstein-Maxwell {\it pp\,}-waves. To date, there are several extensions of
noncommutative gravity~\cite{grav-def,Munich,AG-VM, Mukherjee}, all of them
sharing the property that corrections to the Einstein-Hilbert action start at
order two in $\th^{\a\b}$. Barring thus the eventuality that some specific
symmetries might get lost through the coupling to a SW gauge field, our
results do not depend on the particular model chosen for noncommutative
gravity.

In our solutions, the noncommutativity parameters $\th^{\a\b}$ enter the {\it
  pp\,}-wave metric through a conformal factor $\ho$ that depends on the null
coordinate and which is obtained by solving a linear second order differential
equation. As a result, the distance between points in the same null surface is
modified and grows indefinitely for asymptotic values of the null coordinate.
As concerns the EM field, it receives types of noncommutative corrections: a
susceptibility contribution caused by gravity, a constant
polarization/induction contribution that can be tuned at will and a nonlinear
contribution similar to those in nonlinear optics.

It would also be interesting to investigate the relation of the model and the
solutions presented here with the exact Seiberg-Witten maps studied in
ref.~\cite{Banerjee} and their Dirac-Born-Infeld low energy effective actions.
Another question lying ahead is the generalization to higher dimensions,
especially in relation with compactified extra dimensions, for these may lower
the energy scale at which gravity, and hopefully the noncommutativity scale,
is felt~\cite{extra-dim}.

We note that, so far, noncommutative corrections to conventional {\it
  pp\,}-waves at order one in $\th^{\a\b}$ are not known for pure gravity.
This is in part due to the fact that noncommutative corrections in the
available noncommutative generalizations of general relativity start at order
two~\cite{Munich,Mukherjee}. Technically, the EM sector that we have
introduced provides an order-one source for gravity, which in turn reacts back
modifying the order-one Seiberg-Witten noncommutative correction to the EM
field. The intractability of the order-two corrections to the Einstein-Hilbert
action makes one look for other ways to approach noncommutativity corrections
to general relativity solutions. The result obtained here, namely, that
noncommutativity goes into the metric through a conformal factor, suggests
approaching noncommutative {\it pp\,}-waves in pure gravity by considering the
Seiberg-Witten limit in string scenarios, with the dilaton accounting for
noncommutativity.

\begin{acknowledgments}
Partial support by the BMFB, Germany, under project 05HT4PSA/6 and from CICyT
and UCM-CAM, Spain, through grants~FIS2005-02309,~910770 is acknowledged.
\end{acknowledgments}


\begin{thebibliography}{99}

\bibitem{Doplicher}
   S. Doplicher, K. Fredenhagen and J. E. Roberts, ``The quantum structure of
   spacetime at the Planck scale and quantum fields'',
   Commun. Math. Phys. {\bf 172} (1995) 187.

\bibitem{Szabo}
   R. J. Szabo, ``Symmetry, gravity and noncommutativity'' [arXiv:hep-th/0606233].

\bibitem{grav-def} 
  J. W. Moffat, `` Noncommutative quantum gravity'',
  Phys. Lett. {\bf B491} (2000) 345 [arXiv:hep-th/0007181].\\[3pt]
  A. H. Chamseddine,  Complexified gravity in noncommutative spaces'', 
  Commun. Math. Phys {\bf 218} (2001) 283 [arXiv:hep-th/0005222].\\[3pt]
  H. Garc\'{\i}a-Compe\'an, O. Obreg\'on, C. Ram\'{\i}rez and M. Sabido,
  `` Noncommutative selfdual gravity'', Phys. Rev. {\bf D68} (2003) 044015
  [arXiv:hep-th/0302180]. \\[3pt]
  H. Nishino and S. Rajpoot, ``Teleparallel complex gravity as foundation for
  noncommutative gravity'', Phys. Lett. {\bf B532} (2002) 334
  [arXiv:hep-th/0107216]. \\[3pt]
  D. V. Vassilevich,  ``Quantum noncommutative gravity in two dimensions'',
  Nucl. Phys. {\bf B715} (2005) 695 [arXiv:hep-th/0406163].\\[3pt]
  A. Kobakhidze, ``Theta-twisted gravity'' [arXiv:hep-th/0603132].\\[3pt]
  C. Deliduman, `` Noncommutative gravity in six dimensions''
  [arXiv:hep-th/0607096].\\[3pt]
  E. Harikumar and V. O. Rivelles, `` Noncommutative gravity''
  [arXiv:hep-th/0607115].
  A. H. Chamseddine, ``Applications of the gauge
  principle to gravitational interactions'', Int. J. Geom. Meth. Mod. Phys.
  {\bf 3} (2006) 149 [arXiv:hep-th/0511074]. \\[3pt]
  A. H. Chamseddine, ``Deforming Einstein's gravity'', Phys. Lett. {\bf B504}
  (2001) 33 [arXiv:hep-th/0009153].

\bibitem{Munich}
   P. Aschieri, C. Blohmann, M. Dimitrijevi\'c, F. Meyer, P. Schupp and J. Wess,
   ``A gravity theory on noncommutative spaces'', Class. Quant, Grav. {\bf 22}
   (2005) 3511 [arXiv:hep-th/0504183].\\[3pt]
   P. Aschieri, M. Dimitrijevi\'c, F. Meyer and J. Wess,
   ``Noncommutative geometry and gravity'', Class. Quant, Grav. {\bf 23}
   (2006) 1883 [arXiv:hep-th/0510059].

\bibitem{Seiberg-Witten}
   N. Seiberg and E. Witten, ``String theory and noncommutative geometry'',
   JHEP {\bf 9909} (1999)~032 [arXiv:hep-th/9908142].
   
 \bibitem{AG-VM} 
   L. Alvarez-Gaum\'e, F. Meyer and M. A. V\'azquez-Mozo,
   ``Comments on noncommutative gravity'' [arXiv:hep-th/0605113].
   
 \bibitem{black-hole} 
   P. Nicolini, A. Smailagic and E. Spalucci,
   ``Noncommutative geometry inspired Schwarzschild black hole'', Phys. Lett.
   {\bf B632} (2006)
   547 [arXiv:gr-qc/0510112].\\[3pt]
   T. G. Rizzo, ``Noncommutative inspired black holes in extra dimensions''
   [arXiv:hep-ph/0606051].

\bibitem{Mukherjee}
   P. Mukherjee and A. Saha, ``A note on the noncommutative correction to
   gravity,''  Phys. Rev. {\bf D74} (2006) 027702
   [arXiv:hep-th/0605287].\\[3pt]
   X. Calmet and  A. Kobakhidze, ``Second order noncommutative corrections to
   gravity'' [arXiv:hep-th/0605275].

\bibitem{Guralnik}
   Z. Guralnik, R. Jackiw, S. Y. Pi and A. P. Polychronakos, ``Testing
   non-commutative QED, constructing noncommutative MHD'', Phys. Lett. {\bf
   B517} (2001) 450 [arXiv:hep-th/0106044].

\bibitem{Colladay}
   D. Colladay and V.~A. Kosteleck\'y,
   ``Lorentz-violating extension of the standard model'',
   Phys. Rev. {\bf D58} (1998)~116002 [arXiv:hep-ph/9809521].\\[3pt]
   J. M. Gracia-Bond\'{\i}a, F. Lizzi, F. Ruiz Ruiz and P. Vitale,
   ``Noncommutative spacetime symmetries: Twist versus covariance'',
   Phys. Rev. {\bf D74} (2006) 025014 [arXiv:hep-th/0604206].

\bibitem{Gayral-GB-Ruiz}
  V.~Gayral, J.~M.~Gracia-Bond\'{\i}a and F.~Ruiz Ruiz,
  ``Position-dependent noncommutative products: Classical construction and
  field theory,'' Nucl. Phys. B {\bf B727} (2005) 513 [arXiv:hep-th/0504022].  

\bibitem{Stephani} 
   H. Stephani, D. Kramer, M. Maccallum, C. Hoenselaers and E.
   Herlt, ``Exact solutions of Einstein's field equations'' Cambridge
   University Press (Cambridge 2003).

\bibitem{Ehlers} 
   J. Ehlers and W. Kundt, ``Exacts solutions of the gravitational field
   equations'', in ``Gravitation: an introduction to current research'',
   ed. L. Witten, p. 49, Wiley (New York and London 1962).

\bibitem{Debever}
   R. Debever and M. Cahen, ``Champs \'electromagn\'etiques constants en
   relativit\'e generale'', C. R. Acad. Sci. (Paris) {\bf 251} (1960) 1160. 
  
\bibitem{Abe}
   Y. Abe, R. Banerjee and I. Tsutsui, ``Duality symmetry and plane waves in
   noncommutative electrodynamics'', Phys. Lett. {\bf B573} (2003) 248 
   [arXiv:hep-th/0306272].\\[3pt]
   N. Chatillon and A. Pinzul, ``Light propagation in a background field 
   for time-space noncommutativity and axionic noncommutative QED'',
   [arXiv:hep-th/0608179].
  
\bibitem{Cai}
   R.-G. Cai, ``Superluminal noncommutative photons'', Phys. Lett. {\bf B517}
   (2001) 457 [arXiv:hep-th/0106047].

\bibitem{Banerjee} 
   R. Banerjee and H. S. Yang, `` Exact Seiberg-Witten, induced
   gravity and topological invariants in noncommutative field theories'',
   Nucl. Phys. {\bf B708} (2005) 434 [arXiv:hep-th/0404064].

\bibitem{extra-dim}
   I. Antoniadis, N. Arkani-Hamed, S. Dimopoulos and G. R. Dvali, `` New
   dimensions at a millimeter to a Fermi and superstrings 
   at a TeV'', Phys. Lett. {\bf B436} (1998) 257
   [arXiv:hep-ph/9804398].\\[3pt] 
   L. Randall and R. Sundrum, ``A large mass hierarchy from a small extra
   dimension'', Phys.Rev.Lett. {\bf 83} (1999) 3370 [arXiv:hep-ph/9905221].

\end{thebibliography}
\end{document}